\documentclass[prl,floatfix,twocolumn,showpacs,amsmath,amssymb]{revtex4}
\usepackage{dcolumn}% Align table columns on decimal point
\usepackage{bm}
\usepackage{graphicx}
\usepackage{amsfonts}
\usepackage{amsmath}
\usepackage{amssymb}

\begin{document}

%\twocolumn[\hsize\textwidth\columnwidth\hsize\csname@twocolumnfalse\endcsname

\title{Effective temperature of a dissipative driven mesoscopic system.}
\author{Liliana Arrachea$^{1,2}$ and Leticia F. Cugliandolo$^{2,3}$}
\affiliation{
$^1${Departamento de F\'{\i}sica, Universidad de Buenos Aires, 
Pab. I, Ciudad Universitaria 1428, Argentina}
\\
$^2$Laboratoire de Physique Th{\'e}orique  et Hautes {\'E}nergies,
4 Place Jussieu, 75252 Paris Cedex 05 \\
$^3$Laboratoire de Physique Th{\'e}orique, {\'E}cole Normale
Sup{\'e}rieure,
24 rue Lhomond, 75231 Paris Cedex 05
}

\date{\today}

\begin{abstract}
We study the nonequilibrium dynamics of a mesoscopic metallic ring threaded
by a time-dependent magnetic field and coupled to an electronic reservoir.
We analyze the relation between the (non-stationary) real-time 
Keldysh and retarded Green functions and we find that, in the linear response 
regime with weak heat transfer to the environment, an effective temperature
accounts for the modification of the equilibrium 
fluctuation-dissipation relation.
We discuss possible extensions of this analysis.  
\end{abstract}
\maketitle

Recent technical developments in nanoscience have renewed the interest in 
quantum transport in mesoscopic systems~\cite{1d}. Heat transfer in
these devices is a crucial issue which remains poorly 
understood.

A recent theoretical result in the field of glassy
dynamics is that an `effective temperature', $T_{eff}$, controls
the low-frequency linear response in the limit of small entropy
production (long waiting-time, weak drive)~\cite{Cukupe,Cuku00}.
$T_{eff}$ is defined as the parameter replacing the environmental
temperature, $T$, in the fluctuation-dissipation relation 
({\sc fdr}) evaluated at low-frequencies~\cite{Leto,driven,current}
and it has the properties of a temperature 
in the sense that it
can be measured with a thermometer
and controls heat flows and partial equilibration~\cite{Cukupe,Cuku00}.
A similar phenomenon was found in the relaxation of quantum
spin-glasses~\cite{quantum} and in the Coulomb 
glass~\cite{Grempel}.

The question then arises as to whether the notion of $T_{eff}$ 
plays a role in understanding some features of transport in quantum
mesoscopic systems.  To address it we study the
modification of the {\sc fdr} in a very simple, exactly solvable,
mesoscopic device that consists in a metallic ring 
connected by a lead
to an external particle and thermal reservoir and driven out of
equilibrium by a threading time-dependent magnetic
field~\cite{LB,gefen,yud,Lili,Lili2}.  
In this setting there is an electronic dc current
along the wire and a heat flow towards the
environment.  

We model the metallic wire 
with a system of non-interacting
spin-less electrons described by a one dimensional
periodic tight-binding chain with length $L=Na$ ($N$
is the number of sites and $a$ the lattice spacing),
hopping matrix element $w=W/4$ and    
bandwidth $W$:
\begin{equation}
H^{ring} = - w \sum_{i=1}^N \left( e^{-i \phi t} c_i^\dagger
c_{i+1} + e^{i \phi t} c_{i+1}^\dagger c_i \right) \; .
\label{ring}
\end{equation}
The time-dependent phase $\phi t$ with $\phi\equiv\Phi/(\Phi_0
N)$ and $\Phi_0 = hc/e$ accounts for the external magnetic flux
that we choose to depend linearly in time, $\Phi_M(t) = \Phi t$.
The contact term between the lead and the ring is 
$
H^c=-w_{1\alpha} ( c_1^\dagger c_\alpha + 
c_\alpha^\dagger c_1 )
$.
We model the lead and
reservoir with a 
semi-infinite tight-binding chain with
hopping amplitude $w_\alpha=W_{\alpha}/4$, bandwidth 
$W_\alpha$, and 
spectral density
$\rho_\alpha(\omega)=4 \sqrt{1-\omega^2/W_\alpha^2} 
\; \theta(W_{\alpha}-|\omega|)
$.
We assume that the reservoir is in equilibrium at 
temperature $T$ and chemical potential $\mu$ and  
that its properties are  not
affected by the coupling to the small ring. 
The Hamiltonian of the full system is
then
$H = H^{ring} + H^c + H^{\alpha}$~\cite{Lili}. 
We use a system of units 
such that $\hbar=k_B =\Phi_0 =1$. All our results
have been obtained using a ring with $N=20$ and 
$W=1$, 
and a reservoir with $\mu=-1$ and $W_\alpha=4$.

The dynamics of this problem is amenable to an exact treatment within
the real-time nonequilibrium formalism.  The 
retarded and Keldysh Green functions are 
\begin{eqnarray}
G_{ij}^R(t,t') &\equiv& 
-i \theta(t-t') \langle [c_i(t),c_j^\dagger(t')]_+ \rangle 
\; ,
\label{def:GR}
\\
G^K_{ij}(t,t') &\equiv& 
-i \langle [ c_i(t),c_j^+(t')]_-\rangle 
\label{def:GK}
\; .
\end{eqnarray}
The angular brackets indicate an average computed in the
grand-canonical ensemble, {\it i.e.}  using $H-\mu {\cal N}$ in the
statistical weight with ${\cal N} \equiv \sum_{i=1}^N c_i^\dag c_i$.
The exact evolution equations read~\cite{noneqgenref}
\begin{eqnarray}
& &-i \frac{\partial}{\partial t'}G^R_{ij}(t,t') -  G^R_{ik}(t,t') w_{kj}(t') 
\nonumber\\
& & \;\;\;\;\;\;\;
- \int dt_1 G^R_{ik}(t,t_1) 
\Sigma^R_{kj}(t_1,t') = \delta_{ij} \delta(t-t')\; ,
\label{eq:SDR}\\
& &-i \frac{\partial}{\partial t'}G^K_{ij}(t,t') -  G^K_{ik}(t,t') w_{kj}(t') 
\label{eq:SDK}
\\ 
& & 
= \int dt_1 \left[ G^R_{ik}(t,t_1) \Sigma^K_{kj}(t_1,t') 
+ G^K_{ik}(t,t_1) \Sigma^R_{kj}(t_1,t') 
\right]
\; ,
\nonumber
\end{eqnarray}
with $w_{kj}(t') \equiv w \, e^{i \phi t'} \delta_{k,j+1} +
w \, e^{-i\phi t'} \delta_{k,j-1}$ and the summation convention over 
repeated indices.

We call $g_\alpha^{R,K}$ the Green functions of the bath,
\begin{eqnarray}
g_\alpha^R(\tau) &=& -i \theta(\tau) \int \frac{d\omega}{2\pi} \;
\rho_\alpha^{gc}(\omega) \, e^{-i \omega\tau} 
\; , 
\\ 
g^K_\alpha(\tau)&=&-i
\int \frac{d\omega}{2\pi} \; \tanh(\beta\omega/2) 
\; \rho_\alpha^{gc}(\omega) \, e^{-i \omega\tau}
\; ,
\label{lead}
\end{eqnarray}
$\tau\equiv t-t'$, $\beta=1/T$ and 
$\rho^{gc}_\alpha(\omega)=\rho_\alpha(\omega+\mu)$.
These functions are stationary since the bath is assumed to be 
in equilibrium.
The effect of the
lead and reservoir is then captured by a correction at the
contact site $i=1$~\cite{Pastawski}:
\begin{eqnarray}
\Sigma_{kj}^{R,K}(t,t') = |w_{1\alpha}|^2 g_\alpha^{R,K}(t-t')
\; \delta_{k1} \delta_{j1} \; .
\label{bath}
\end{eqnarray}

If one assumes that the magnetic field and coupling to the reservoir
had been switched on in the far past in such a way that any transient
behaviour dependent on the initial conditions died out at the working
times $(t,t')$,
Eq.~(\ref{eq:SDK}) reduces to
\begin{displaymath}
G^K_{ij}(t,t')= \int dt_1 dt_2 \; G^R_{ik}(t,t_1)
\Sigma^K_{kl}(t_1,t_2) G^A_{lj}(t_2,t')
\label{intgk}
\end{displaymath}  
that yields $G_{ij}^K$ 
once $G_{ij}^R$ has been computed from Eq.~(\ref{eq:SDR})~\cite{Lili}.

In equilibrium a model-independent fluctuation-dissipation 
relation ({\sc fdr}) between the retarded and 
Keldysh Green functions holds. 
Calling $G_{ij}^{R,K}(t,\omega)$ 
the Fourier transform with respect to  
$\tau$ the {\sc fdr} reads
\begin{eqnarray}
 G^{K}_{ij}(t,\omega)  =
\tanh(\beta\omega/2)
\left( G_{ij}^{R}(t,\omega) - G_{ij}^{A}(t,\omega)\right) 
\label{fermionic-fdt-grancan}
\end{eqnarray} 
where, for later convenience, we kept explicit a dependence on the
observation time $t$ that does not exist in equilibrium.  Out of
equilibrium there is no reason why such a relation should hold, and
the temperature and chemical potential of the system are not even
defined. Our aim is to determine how the relation
(\ref{fermionic-fdt-grancan}) should be modified to account for the
Green functions of the ring.

\begin{figure}[t]
\includegraphics[width=80mm,angle=-90]{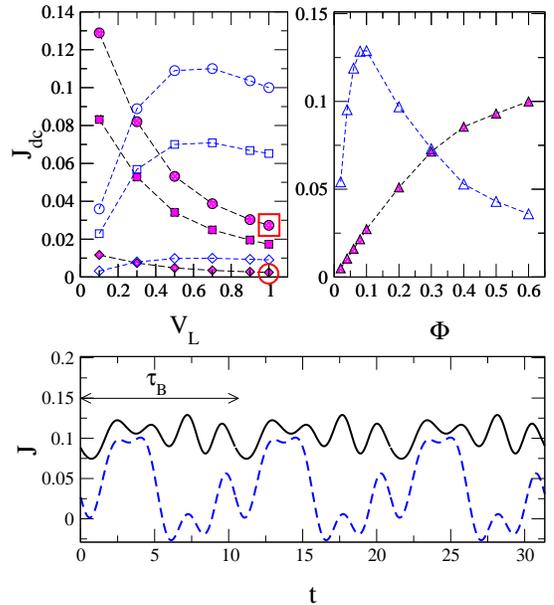}
\caption{(Color online.) Dependence of the current on time and various
parameters. Upper-left panel: the $dc$ current $J_{dc}$ against the
coupling to the lead $V_L$ for $\Phi=0.6$ (open blue symbols) and
$\Phi=0.1$ (filled magenta symbols).  Different symbols correspond to
$T=0.05$ (circles), $T=1$
(squares) and $T=10$ (diamonds).  The large circle and square indicate
the values of $J_{dc}$ that correspond to the data shown in
Figs.~\ref{Fig2}, \ref{Fig3} and \ref{Fig4}.  Upper-right panel:
$J_{dc}$ against $\Phi$ for $T=0.05$.  Open symbols with dashed blue
line and solid symbols with dashed black line correspond to $V_L=0.1$
and $V_L=1$, respectively.  Lower panel: $J_{12}(t)$ for $\Phi=0.6$,
$T=0.05$, $V_L=0.1$ (solid black line) and $V_L=0.9$ (dashed blue
line). The period $\tau_B \approx 10$ is indicated with an arrow.}
\label{fig:uno}
\end{figure}

First, let us discuss the values of the parameters $T$, $\Phi$ and
$V_L=|w_{1\alpha}|^2$ for which we may expect a simple generalization
of the {\sc fdr}. To this end we summarize the dependence of the
current on time and these parameters in
Fig.~\ref{fig:uno}~\cite{Lili,Lili2}.  In the lower panel we
show the time-dependence of the current $J_{l,l+1}(t) \equiv 2ei \,
\mbox{Re}[w e^{-i \phi t} \langle c^\dagger_l(t) c_{l+1}(t) \rangle]$
for $l=1$, $\Phi=0.6$, $T=0.05$ and two values of the coupling to the
lead, $V_L=0.9$ (solid black line) and $V_L=0.1$ (dashed blue
line). The current has the period $\tau_B=2\pi/\Phi$ of Bloch
oscillations.  Their amplitude is diminished for large enough $T$ and
$V_L$ and for small enough $\Phi$ as well as for very large $\Phi (>
W)$.

Dissipation gives rise to a dc-component in the current,
$J_{dc}=\tau_B^{-1} \int_0^{\tau_B} dt\; J_{l,l+1}(t)$, that is
related to the power at which energy is dissipated in the form of heat
through $P=J_{dc} \Phi$ as has been verified exactly~\cite{Lili2}.  In
the upper-left panel in Fig.~\ref{fig:uno} we show the dependence of
$J_{dc}$ on the strength of the coupling to the reservoir, $V_L$, for
several values of the electromotive force (emf), $\Phi$, and
temperature.  At fixed temperature $J_{dc}$ is a nonmonotonic function
of $V_L$ with the boundary values $J_{dc}(0)=J_{dc}(\infty)=0$
and $J_{dc} \propto V_L$ when $V_L \rightarrow 0$ \cite{Lili,leh}.  For
rather large values of the bias ($\Phi=0.6$ open symbols) the maximum
occurs at $V_L \approx 0.5$.  For small values of the emf ($\Phi=0.1$
filled symbols) the first increasing part of the curve is squeezed
towards very low values of $V_L$ and only the second decreasing regime
is visible.  For fixed $\Phi$, the higher $T$ the lower $J_{dc}$, as
can be checked by comparing the curves with different symbols.

The upper-right panel in Fig.~\ref{fig:uno} 
shows that $J_{dc}$ is in approximate linear
relation with $\Phi$ for small biases; deviations appear at a value
$\Phi_m$ that increases with $V_L$, {\it e.g.}
$\Phi_m\approx 0.075$ for $V_L=0.1$ and $\Phi_m\approx 0.2$ for
$V_L=1$.
 
\begin{figure}[t]
\includegraphics[width=65mm,angle=-90]{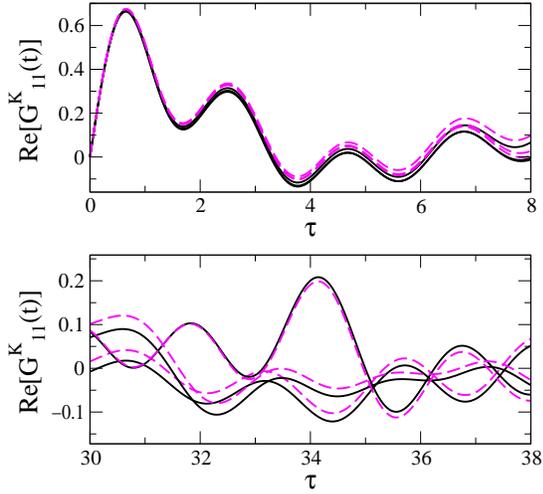}
\caption{(Color online.) Breakdown of stationarity and the {\sc fdr}.
Real part of $G^K_{11}(t,\tau)$ (solid black lines) and the inverse
Fourier transform of the rhs of Eq.~(\ref{fermionic-fdt-grancan})
(dashed magenta lines) as a function of $\tau$ for three total times
$t$ equally spaced in the interval $[0,\tau_B]$. $V_L=1$, $\Phi=0.1$,
$T=0.05$ and $J_{dc}\approx 0.03$ (red square in Fig.~\ref{fig:uno}).
}
\label{Fig2}
\end{figure}

We now turn to the detailed study of the Green functions that we
parametrize as $G_{ij}^{R,K}(t,t-\tau)$. In Fig.~\ref{Fig2} we
illustrate the non-trivial dependence on $t$ and $\tau \equiv t-t'$ by
tracing, with solid black lines, Re$G_{ij}^K(\tau)$ for three values
of the total time $t$ equally spaced within $\tau_B$.  The two panels
show the evolution on the intervals $0 \leq \tau \leq 8$ (upper) and
$30 \leq \tau \leq 38$ (lower).  Interestingly enough, all curves fall
on top of each other for $\tau \leq \tau^*(\Phi,\beta,V_L) \approx
1.5$, and they later deviate demonstrating the breakdown of
stationarity.  This trend can be understood by noting that
$G^{R}_{i,j}(t,\omega)$ contains a time dependent structure within the
spectral range of the free ring, {\it i.e.} in the interval $|\omega +
\mu| \leq W/2$, while the high frequency part is dominated by the {\em
stationary} spectral features of the reservoir which set the quick
response in the time domain.

Figure~\ref{Fig2} also shows a qualitative study of the {\sc fdr}.
The dashed lines are the inverse Fourier transform of the rhs of
Eq.~(\ref{fermionic-fdt-grancan}) for the same values of $t$.  The
companion curves do not match and the {\sc fdr} does not
hold.  We note, however, that for $\tau \leq \tau^*\approx 1.5$ the
behaviour is not only stationary but the {\sc fdr} holds as well.

\begin{figure}[t] 
\includegraphics[width=65mm,angle=-90]{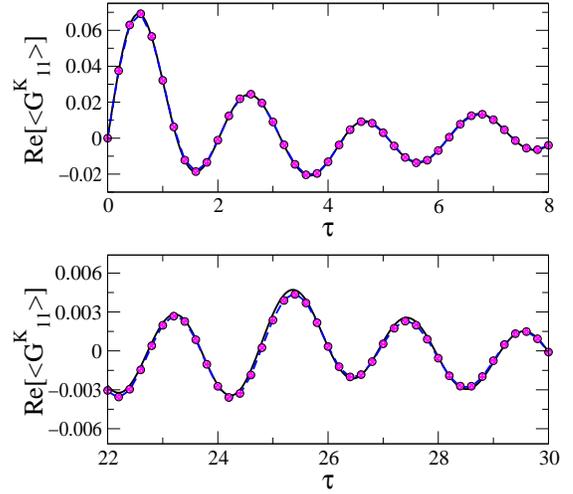}
\caption{(Color online.) Study of the {\sc fdr} in the `classical' regime,
$T=10$.  Real part of $\langle G^K_{11}\rangle (\tau)$ (dashed blue
line) and the inverse Fourier transform of the rhs of
Eq.~(\ref{fermionic-fdt-grancan}) (solid black line) averaged over a
period $\tau_B$, see Eq.~(\ref{averaged-Gf}), as a function of
$\tau$. The two curves are almost identical, apart from some little
deviations seen, {\it e.g.}, at $\tau \approx 25.5$.  The symbols are
obtained using the same functional form in the rhs of
Eq.~(\ref{fermionic-fdt-grancan}) with $\beta$ replaced by a fitting
parameter $T_{eff}=10.02\approx T$.  $V_L=1$, $\Phi=0.1$,
$J_{dc}\approx 0.002$ (red circle in Fig.~\ref{fig:uno}).}
\label{Fig3} 
\end{figure}

As Fig.~\ref{Fig2} demonstrates, the total time dependence is very
complicated. Instead of studying the modification of the {\sc fdr} for
each value of $t$ we found it natural to work
with the averaged Green functions:
\begin{equation}
\langle G_{ij}^{K,R}\rangle (\tau)  \equiv \frac{1}{\tau_B} 
\int_0^{\tau_B} dt \; 
G_{ij}^{K,R}(t,\tau)
\; .
\label{averaged-Gf}  
\end{equation}
In Fig.~\ref{Fig3} we show a quantitative test of the {\sc fdr} in a
`classical' regime (loosely) identified as the values of the
environmental temperature $T$ such that the classical limit of the
{\sc fdr}, $g_\alpha^K(\tau)=-i/(2T) \partial_{\tau}
g^R_{\alpha}(\tau)$, $\tau \geq 0$, holds for the bath.  We compare
the left and right hand sides of the inverse Fourier transform of
Eq.~(\ref{fermionic-fdt-grancan}) averaged over $t$ as defined in
Eq.~(\ref{averaged-Gf}).  In both intervals shown the accord between
the two curves is rather good, proving that the {\sc fdr}
approximately holds when the driving force is weak and the temperature
of the environment is high in such a way that $P$ is extremely low.

\begin{figure}[t]
\includegraphics[width=65mm,angle=-90]{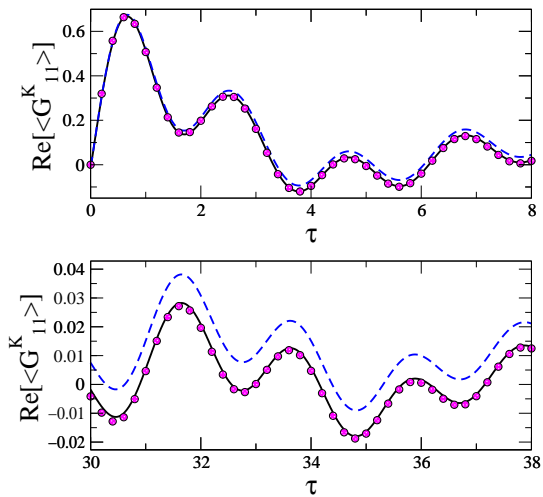}
\caption{(Color online) {\sc fdr} in the `quantum' regime, $T=0.05$.
$\mbox{Re}\langle G^K_{11}\rangle$ (dashed blue line) and inverse
Fourier transform of the rhs of Eq.~(\ref{fermionic-fdt-grancan})
(solid black line) averaged over a period $\tau_B$ against $\tau$.
The symbols correspond to the latter functional form with a fitting
parameter $T_{eff}=0.143$. $\Phi=0.1$, $V_L=1$ and $J_{dc}\approx
0.03$ (red square in Fig.~\ref{fig:uno}).}
\label{Fig4} 
\end{figure}

In Fig.~\ref{Fig4} we test the {\sc fdr} in the `quantum' regime,
$T=0.05$.  The symbols represent a fit using the functional form in
the rhs of Eq.~(\ref{fermionic-fdt-grancan}) with $T$ replaced by
$T_{eff}=0.143$.  For short time-differences,
say before the first minimum in the upper panel, the {\sc fdr} holds --
and the fit also falls on top of the original curves. Indeed, 
if $\beta_{\sc eff}\omega \gg 1$, 
$\tanh(\beta_{eff}\omega/2) \approx 1$ and 
the Keldysh and retarded Green functions are in linear relation. 
For longer time
differences the deviations are clear but the fit
accounts well for the data.

Within the accuracy of our numerical solution, the effective
temperature describes rather correctly the modification of the {\sc
fdr} whenever the dc-current is in linear relation with the emf $\Phi$
(see Fig.~\ref{fig:uno}).  Out of the linear response regime one
cannot match the two sides of Eq.~(\ref{fermionic-fdt-grancan}) by
simply using a single-valued $T_{eff}$. This limitation parallels the
one observed in classical {\it strongly driven} glassy
systems~\cite{driven,current}.  This is reasonable since the notion of
an effective temperature was proposed to apply to a regime of small
entropy production only~\cite{Cukupe,Cuku00} and this, in our case,
corresponds to low $P=J_{dc}\Phi$ values.

\begin{table}[h]
\begin{tabular}{ccccccc}
\tableline
\tableline
$V_L\;\;\;\;\;\;$ & $\Phi\;\;\;\;\;\;$ & $T\;\;\;\;\;\;$ & 
$T_{eff}\;\;\;\;\;\;$ & $(T_{eff}-T)/T$ & $P (\times 10^{-3})$ \\
\tableline
1        & 0.1    & 0.05  & 0.143   &1.84            & 2.7       \\
1        & 0.1    & 0.1   & 0.167   & 0.67           & 2.6       \\
1        & 0.1    & 0.2   & 0.264     & 0.32           & 2.5     \\
1        & 0.1    & 1     &  1.052  & 0.052          & 1.7        \\
1        & 0.1    & 10    &  10.02  & 0.002          & 0.3       \\
\tableline
0.1        & 0.06    & 0.05  & 0.769  &14.38         & 7.1       \\
0.1        & 0.06    & 0.1   & 0.8    & 7            & 7.0       \\
0.1        & 0.06    & 0.2   & 0.833  & 3.16         & 6.9     \\
0.1        & 0.06    & 1     &  1.43  & 0.43          & 4.6        \\
\tableline
\end{tabular}
\caption{Estimates of the effective temperature and dissipated power
within the linear response regime.}
\end{table}

In all cases $T_{eff}> T$, as also found in glassy systems relaxing
from a disordered initial configuration and in driven classical
systems~\cite{Leto}.  An heuristic argumement that explains this
result is the following.  The equations that govern the dynamics of
the Hamiltonian $H$ are the same as those of an effective Hamiltonian
where the electrons are coupled to modes with energy quanta $\Phi$
\cite{Lili2}. It is natural to expect that such a coupling enables a
mechanism for the electrons to spread out in energy, which can be
interpreted as an effective increment of the electronic temperature.
More quantitatively, we found that $T_{eff}$ increases with the power
dissipated, $P=J_{dc} \Phi$. Some rough estimates are shown in Table~1.

Let us discuss 
some open questions.  So far $T_{eff}$ is a
fitting parameter with a meaningful physical interpretation.  It
remains to be proven~\cite{Cukupe,Cuku00}, whether we can associate 
it to a {\it
%, as done for classical glassy problems
bonafide} temperature for nonequilibrium quantum systems with
fermionic statistics.

We did not need to use a $\mu_{eff}$ since $\mu$ basically controls
the period of the oscillations of the inverse Fourier transform of the
{\sc rhs} of Eq.~(\ref{fermionic-fdt-grancan}) that coincides with the
period of $\mbox{Re} \langle G_{11}^K\rangle (\tau)$, see
Figs.~\ref{Fig3} and \ref{Fig4}. In other nonequilibrium settings an
upgrading of $\mu$ to $\mu_{eff}$ might be necessary.

We defined $T_{eff}$ via the Green functions that are not, however,
directly accessible experimentally.  A similar description should
apply to the current fluctuations~\cite{current} that can be measured
in noise experiments~\cite{Imry}.

Our problem bears
some resemblance with a {\it classical ratchet} for which a $T_{eff}$
could be defined in the strongly driven limit for long wave-length
observables~\cite{Sasa}. It would be interesting to check whether 
this holds for quantum mesoscopic devices too.
Even if we have not found a site-dependent $T_{eff}$ for 
weak forcings we cannot exclude such a dependence
in a more general setting. 

Finally, it would be interesting to check 
the modifications of the {\sc fdr} using 
different time-dependent magnetic fields, including disorder in the
hopping rates, considering different dissipative mechanisms, etc. 

We thank SISSA, MPIPKS-Dresden and A. von
Humboldt Stiftung  (LA), ICTP and UBA (LFC)
for hospitality. We acknowledge financial support from Ecos-Sud
and ACI grants
(LFC), CONICET and PICT 03-11609 (LA).  
We thank C. Chamon, B. Doucot, J. Kurchan, G. Lozano
 and Prof. Fulde for very helpful discussions.

\end{document}